\documentclass[lettersize,journal]{IEEEtran}

\IEEEoverridecommandlockouts

\usepackage{amsfonts}
\usepackage[dvips]{graphicx}
\usepackage{times}
\usepackage{cite}
\usepackage{amsmath}
\usepackage{cases}
\usepackage{array}
\usepackage{dsfont}
\usepackage{amssymb}

\usepackage{siunitx}
\usepackage{caption}

\usepackage{stfloats}

\usepackage{graphicx}
\usepackage{subfigure}
\usepackage{pdfpages}
\usepackage{footnote}
\usepackage{booktabs}
\usepackage{soul}
\usepackage{array}
\usepackage{multirow}
\usepackage{bm}
\usepackage{empheq}
\usepackage{amsthm}
\usepackage{algorithm}
\usepackage{algorithmicx}
\usepackage{algpseudocode}
\usepackage{color}
\usepackage{diagbox}
\usepackage{caption}
\theoremstyle{plain}

\theoremstyle{definition}

\usepackage[colorlinks,bookmarksopen,bookmarksnumbered,citecolor=blue, linkcolor=blue, urlcolor=blue]{hyperref}

\let\oldReturn\Return
\renewcommand{\Return}{\State\oldReturn}

\usepackage{textcomp}
\usepackage{xcolor}

\ifCLASSINFOpdf
\else
\fi

\begin{document}
	\title{Joint Source-Channel Noise Adding with Adaptive Denoising for Diffusion-Based Semantic Communications}
	\vspace{-4pt}
	
	\vspace{-25pt}\author{\IEEEauthorblockN{$\text{Chengyang Liang}$, $\text{Dong Li}, ~\IEEEmembership{Senior Member,~IEEE}$}
		\vspace{-22pt}
		\thanks{Chengyang Liang and Dong Li are with the School of Computer Science and Engineering, Macau University of Science and Technology, Macau, China (e-mail: 3240006992@student.must.edu.mo; dli@must.edu.mo).}}    
	\vspace{-20pt}

\maketitle
\begin{abstract}
	Semantic communication (SemCom) aims to convey the intended meaning of messages rather than merely transmitting bits, thereby offering greater efficiency and robustness, particularly in resource-constrained or noisy environments. In this paper, we propose a novel framework which is referred to as joint source-channel noise adding with adaptive denoising (JSCNA-AD) for SemCom based on a diffusion model (DM). Unlike conventional encoder-decoder designs, our approach intentionally incorporates the channel noise during transmission, effectively transforming the harmful channel noise into a constructive component of the diffusion-based semantic reconstruction process. Besides, we introduce an attention-based adaptive denoising mechanism, in which transmitted images are divided into multiple regions, and the number of denoising steps is dynamically allocated based on the semantic importance of each region. This design effectively balances the reception quality and the inference latency by prioritizing the critical semantic information. Extensive experiments demonstrate that our method significantly outperforms existing SemCom schemes under various noise conditions, underscoring the potential of diffusion-based models in next-generation communication systems.

\end{abstract}

\begin{IEEEkeywords}Semantic communication, Joint source-channel noise adding, diffusion model, adaptive denoising, attention mechanism.
\end{IEEEkeywords}

\IEEEpeerreviewmaketitle

\vspace{-4pt}
\section{Introduction} 
\label{sec:Introduction}

\IEEEPARstart{T}{he} explosive growth of data traffic, fueled by applications such as the Internet of Things (IoT), autonomous systems, and immersive media, is pushing conventional communication systems, designed under the Shannon paradigm, to their limits. Traditional methods primarily focus on maximizing data rates and minimizing bit errors, often neglecting the semantic meaning of transmitted data. However, emerging applications prioritize successful interpretation over bit-perfect recovery \cite{AComplete, SCcarrier}, motivating the development of Semantic Communication (SemCom) framework which emphasizes task-relevant information transmission, and promises to reshape future wireless networks.

Initially conceptualized decades ago, SemCom has recently advanced with deep learning (DL) breakthroughs. Contemporary DL-based approaches often adopt Joint Source-Channel Coding (JSCC), which integrates source compression and channel protection within unified deep neural networks \cite{DeepJSCC, NLJSCNSC}. Nevertheless, these systems face challenges in accurately reconstructing complex semantics, particularly for high-dimensional data such as images, and often experience performance degradation under low Signal-to-Noise Ratio (SNR) conditions, hindering the practical deployment.

Denoising Diffusion Probabilistic Models (DDPMs) have emerged as a powerful class of deep generative models \cite{DDPM}. DDPMs function through two primary processes: a forward noising procedure that gradually perturbs data towards a simple noise distribution, and a learned reverse denoising process that iteratively refines noise into realistic data samples. The advantage is that they can achieve the state-of-the-art performance in generating high-fidelity and diverse outputs across domains such as image synthesis, often exhibiting more stable training and superior sample quality compared to Generative Adversarial Networks (GANs) and Variational Autoencoders (VAEs). The intrinsic ability of the diffusion model (DM) to manage the noise and their strong generative capacity suggest considerable potential for enhancing SemCom systems, particularly in noisy and resource-constrained environments. Recent preliminary studies have explored the application of DMs in related areas such as semantic compression and communication \cite{YOLO, SelectSC}, demonstrating promising results. 

However, current approaches primarily treat communication noise as an external disturbance to be mitigated, without fully exploiting the unique structure of the diffusion process. Specifically, the challenge of seamlessly integrating channel impairments into the diffusion framework, leveraging noise as a constructive element rather than a harmful factor, remains an open challenge. Moreover, adaptive strategies that dynamically allocate denoising resources based on semantic importance are still underexplored on how to reduce the latency caused by the inference time. These gaps motivate us to rethink the roles of the channel noise and denoising in SemCom, leading to the development of a novel framework termed joint source-channel noise adding with adaptive denoising (JSCNA-AD) for SemCom systems. The contributions of this work are summarized as follows:

\begin{itemize}
	\item \textbf{JSCNA-AD SemCom Framework:} We propose a novel SemCom framework, termed as the JSCNA-AD for SemCom systems, which unifies the source and channel noise through a controlled diffusion-based perturbation process at the encoder, and employs an adaptive attention-guided reverse diffusion strategy at the decoder for region-wise semantic refinement. This joint design transforms the channel noise into a generative prior while dynamically adjusting the denoising effort based on the semantic importance.
	\item \textbf{Joint Source and Channel Noising Adding:} We design a semantic encoder that seamlessly integrates the forward diffusion process with channel-aware noise modeling by injecting both structured  and unstructured noise into the source signal. To align the noisy representation with the generative structure of diffusion models, the encoder incorporates a pre-compensation step using the inverse channel matrix, effectively neutralizing fading effects. This unified perturbation mechanism transforms the traditionally harmful channel noise into a learnable component of the semantic embedding, enabling the transmitter to generate a resilient latent representation that naturally fits the denoising dynamics at the receiver.
	\item \textbf{Adaptive Denoising:} We propose an attention-based adaptive denoising mechanism at the receiver that partitions the received image into semantically meaningful regions and dynamically allocates the number of reverse diffusion steps to each region based on its estimated semantic importance. By prioritizing critical regions with more detailed denoising iterations and reducing the effort on less relevant areas, the decoder achieves a balanced trade-off between the semantic fidelity and the computational latency. This design enables selective refinement of key contents, ensuring robust semantic recovery under varying noise levels while maintaining denoising efficiency.
	\item Extensive simulations demonstrate that our method outperforms the latest SemCom schemes, achieving up to a 13.3\,dB higher PSNR under harsh channel conditions while reducing the inference time by approximately 6\%.
	
\end{itemize}

\vspace{-8pt}
\section{System Model}
\label{sec:system model}

\begin{figure}[t]
\vspace{-2.5mm}
\centering
\includegraphics[scale=0.29]{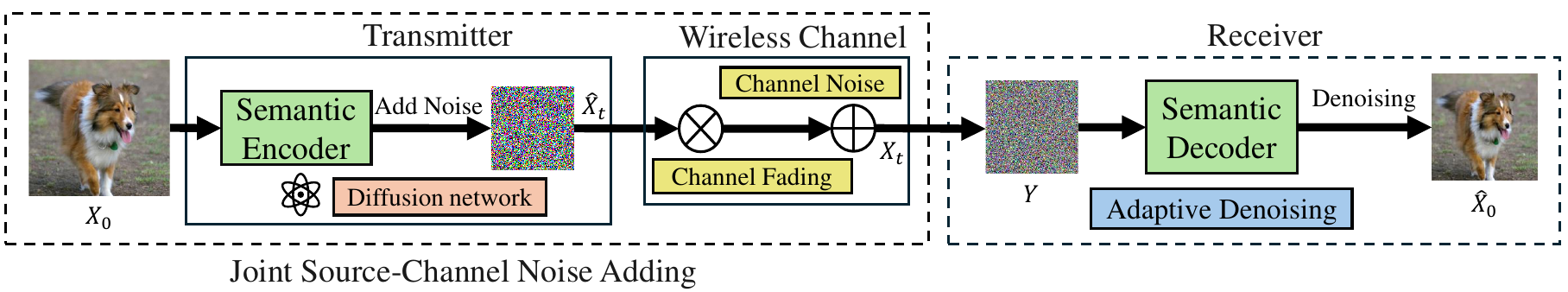}
\vspace{-6pt}
\caption{Illustration of the proposed JSCNA-AD semantic communication system.}
\label{fig:system}
\vspace{-8pt}
\end{figure}

In this section, we design the overall structure of the proposed JSCNA-AD for SemCom system based on DM, as illustrated in Fig. \ref{fig:system}. The system comprises three primary components: 1. a semantic encoder that incorporates the channel noise with inversed channel fading into the forward diffusion process, and 2. a semantic decoder that performs reverse diffusion denoising.

We define the image $X_0$ of the source information as an RGB image signal $X_0 \in \mathbb{R}^{h\times w\times3}=\mathbb{R}^d$, where h is the image height and w is the image width. To facilitate robust transmission over noisy channels, the transmitter incorporates the channel noise to $X_0$, in accordance with the forward diffusion dynamics

\vspace{-2.5mm}
\begin{equation}
	\vspace{-1.5mm}
	X_t=SemE_\phi^{\mathrm{d}}(X_0,t,\mathbf{H}^{-1}),\quad t\in\{1,\ldots,T\},
\end{equation}

\noindent where $SemE_{\phi}^{\mathrm{d}}(\cdot)$ denotes the joint semantic encoder parameterized by $\phi$, incorporating both controlled noise injection and channel-aware transformation, and $\mathbf{H}^{-1}$ denotes the inverse channel.

The transmission medium is modeled as a wireless channel that is affected by fading and additive noise. The received signal $Y\in\mathbb{R}^d$ is represented as follows

\vspace{-1.5mm}
\begin{equation}
	\vspace{-1.5mm}
	Y = \mathbf{H} X_t + n,
\end{equation}

\noindent where $\mathbf{H}\in\mathbb{R}^{d\times d}$ denotes the channel gain matrix, and $n$ represents the channel noise. The overall noise affecting the signal thus combines both the noise in the diffusion forward process and the unintentional channel noise $n$, which are treated in a unified manner during denoising.

At the receiver, the received signal $Y$ directly corresponds to the noisy latent representation $X_t$ corrupted only by channel noise. The decoder then applies the reverse diffusion process $SemD_\rho^{\mathrm{d}}(\cdot)$ to iteratively refine the semantic content

\vspace{-2.5mm}
\begin{equation}
	\vspace{-1.5mm}
\hat{X}_0=SemD_\rho^{\mathrm{d}}(Y,t),
\end{equation}

\noindent where $\rho$ denotes the parameters of the decoder, and the reverse diffusion process iteratively denoises $\hat{X}_{t}$ over multiple steps, utilizing the learned semantic prior to recover $\hat{X}_0\approx X_0$.

\vspace{-6pt}
\section{architecture of encoder and decoder}
\label{sec:architecture}

In this section, we elaborate on the architectural design of the proposed JSCNA-AD framework. 

\vspace{-8pt}
\subsection{Diffusion Network of JSCNA-AD Framework}
\label{sec:Network}

\begin{figure}[tb]
	\vspace{-8pt}
	\centering
	\includegraphics[scale=0.26]{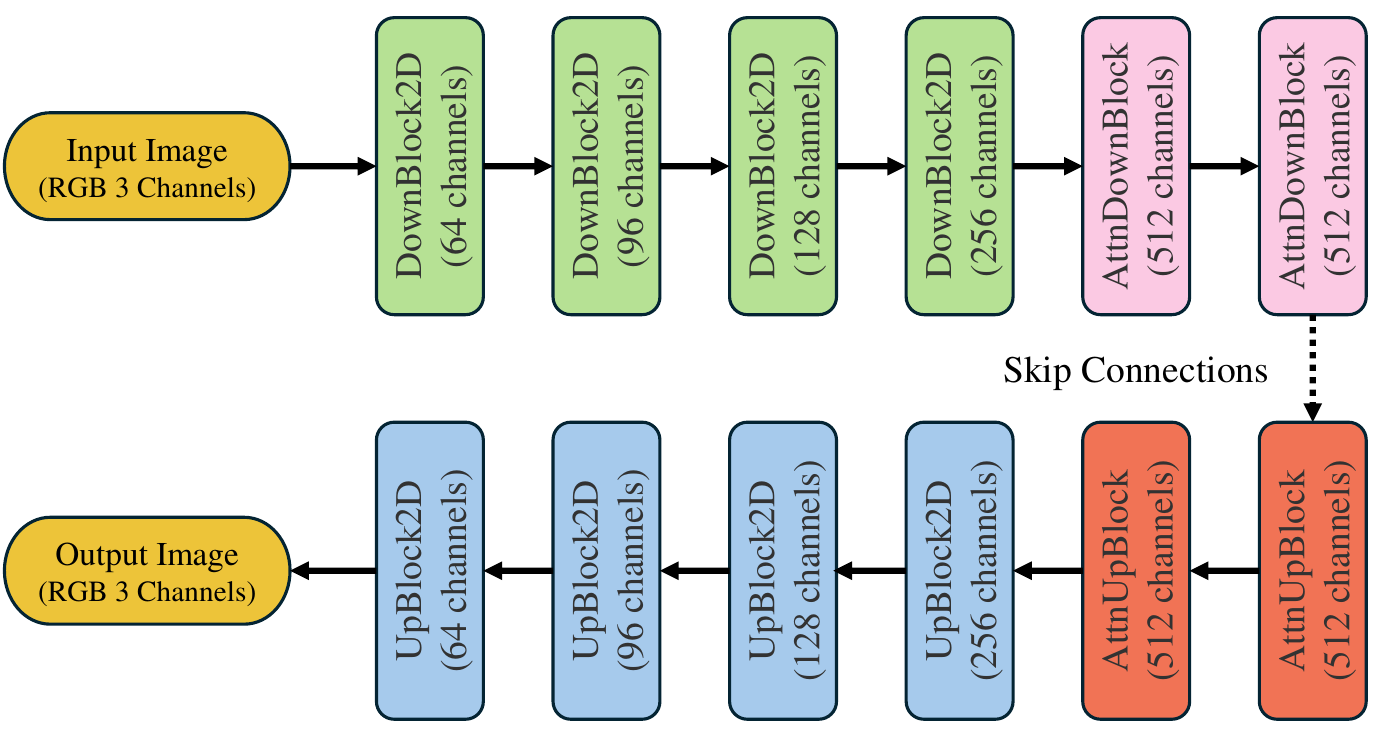}
	\caption{Demonstration of the diffusion network architecture designed for JSCNA-AD framework.}
	\label{fig:unetmodel}
	\vspace{-14pt}
\end{figure}

In our diffusion-based SemCom system, both the forward noise adding process and the reverse denoising process are modeled using a UNet-based neural network, shared by both the transmitter and receiver. As shown in Fig.~\ref{fig:unetmodel}, the architecture comprises six downsampling and six upsampling blocks, with self-attention mechanisms embedded in the two deepest layers to capture long-range semantic dependencies critical for reconstructing structured data.

The downsampling blocks progressively extract abstract features, while upsampling layers refine spatial details. The channel dimensions increase from 64 to 512, stabilizing training under large timestep schedules. Skip connections and attention bottlenecks enhance robustness to channel degradation and semantic variability.

The training of the diffusion-based SemCom model encompasses two primary objectives: estimating the noise introduced during the forward diffusion process and ensuring that the reconstructed semantic representation accurately reflects the original source.

The noise estimation loss is given by
\vspace{-4pt}
\begin{equation}
	\vspace{-2pt}
	\mathcal{L}_{\mathrm{denoise}}=\mathbb{E}_{\mathbf{X}_0,t,\boldsymbol{\epsilon}}\left[\left\|\boldsymbol{\epsilon}-\mu_\Phi\left(X_t,t\right)\right\|_2^2\right],
\end{equation}

\noindent where we define a reconstruction loss between the final output $\hat{X}_0=SemD_\rho^{\mathrm{d}}(X_t,t)$ and the ground truth $X_0$ by employing a hybrid of $\ell_{1}$ and $\ell_{2}$ distances

\vspace{-6pt}
\begin{equation}
	\begin{aligned}
	\mathcal{L}_{\mathrm{rec}}W&=\frac{1}{2}\left\|SemD_\rho^{\mathrm{d}}(X_t,t)-X_0\right\|_1\\
	&+\frac{1}{2}\left\|SemD_\rho^{\mathrm{d}}(X_t,t)-X_0\right\|_2^2.
	\end{aligned}
\end{equation}

The total loss function used during training becomes

\vspace{-8pt}
\begin{equation}
	\vspace{-4pt}
	\mathcal{L}_{\mathrm{total}}=\mathcal{L}_{\mathrm{denoise}}+\beta\cdot\mathcal{L}_{\mathrm{rec}},
\end{equation}

\noindent where $\beta$ is a trade-off coefficient. In practice, this dual-objective loss function enables the model to capture both the pixel-wise detail and the perceptual quality, thereby enhancing its performance across both high and low SNR regimes.

\vspace{-14pt}
\subsection{Joint Source and Channel Noise Adding}
\vspace{-4pt}

The encoder performs joint source compression and channel adaptation by introducing structured noise into the system. The initial state, denoted as $X_0$, is generated through a parameterized forward diffusion process, normalized and aligned to match the statistical characteristics of the wireless channel. The process is mathematically represented as follows

\vspace{-2mm}
\begin{equation}
	\vspace{-0.5mm}
	\tilde{X}_t = \sqrt{\bar{\alpha}_t} X_0 + \sqrt{1 - \bar{\alpha}_t} \cdot \boldsymbol{\epsilon}, \quad \boldsymbol{\epsilon} \sim \mathcal{N}(0, \mathbf{I}),
\end{equation}

\noindent where $\bar{\alpha}_t=\prod_{s=1}^t\alpha_s$ is the cumulative noise schedule, and $\epsilon$ denotes the artificially injected noise during the forward diffusion process. $\tilde{X}_t = \frac{\tilde{X}_t }{\sqrt{\mathbb{E}[|\tilde{X}_t|^2]}}$ this normalization enables the structured noise to be matched with the SNR assumptions of the transmission channel.

Specifically, this process integrates the forward diffusion dynamics with channel pre-compensation, unifying the semantic embedding with physical-layer robustness. The internal mechanism can be decomposed as $X_t =  \mathbf{H}^{-1} \tilde{X}_t $, where $\mathbf{H}^{-1}$ denotes the inverse channel. This formulation ensures that noise injection is an integral part of the generative prior modeling and channel adaptation, jointly forming a resilient latent representation for wireless transmission.

The wireless channel is modeled with fading and additive Gaussian noise. The received signal is given by

\vspace{-1.5mm}
\begin{equation}
	\vspace{-1mm}
	Y = \mathbf{H} X_t + n = \tilde{X}_t + n,
\end{equation}

\noindent where $n\sim\mathcal{N}(0,\sigma_n^2 \mathbf{I})$ denotes the uncontrollable channel noise. This formulation unifies the injected forward diffusion noise and channel noise in a semantically structured latent space.

At the receiver, the observed signal $Y$ is a noisy version of the latent representation. The decoder seeks to iteratively reconstruct the clean semantic signal $\hat{X}_0$ using a reverse diffusion process conditioned on the channel-distorted input. The denoising step is defined as

\vspace{-2.5mm}
\begin{equation}
	\vspace{-0.5mm}
	X_{t-1} = \frac{1}{\sqrt{\alpha_t}} \left( Y - \frac{1-\alpha_t}{\sqrt{1-\bar{\alpha}t}} \boldsymbol{\epsilon}_{\theta}(Y, t) \right) + \sigma_t \mathbf{z}, \quad \mathbf{z}\sim\mathcal{N}(0, \mathbf{I}),
\end{equation}

\noindent where $\boldsymbol{\epsilon}_{\theta}(\cdot)$ is the learned noise estimator and $\sigma_{t}$ is the variance at step $t$. After $T$ steps, we obtain the semantic reconstruction $\hat{X}_{0}$.

\vspace{-6pt}
\subsection{Attention Mechanism-Driven Adaptive Denoising}
\label{sec:attention}
\begin{figure}[t]
	\vspace{-1.5mm}
	\centering
	\includegraphics[scale=0.27]{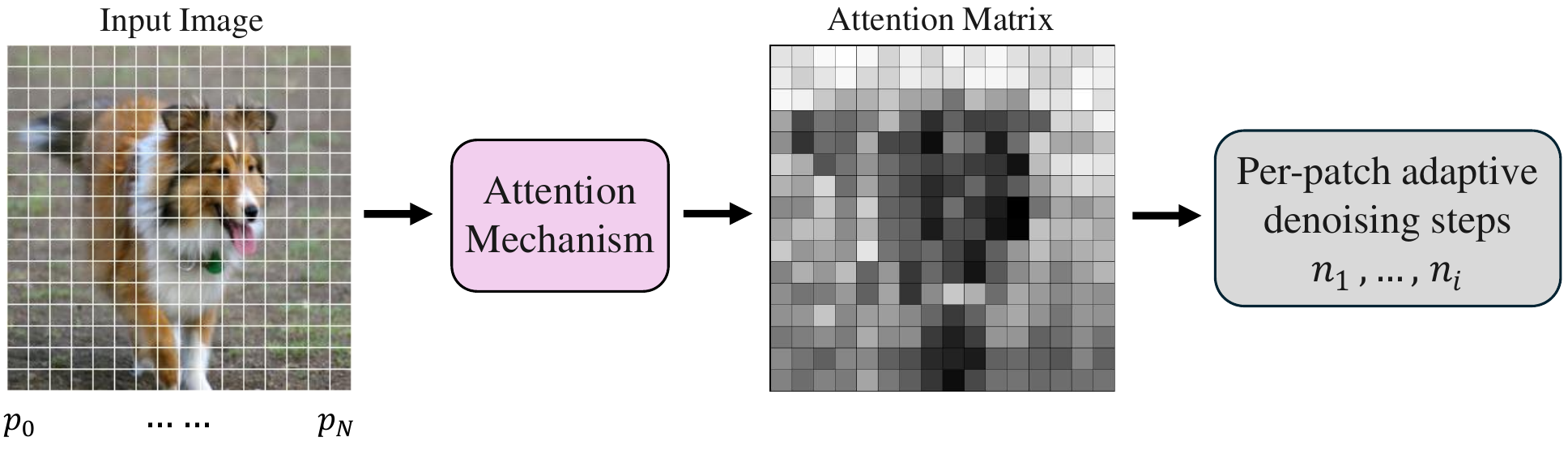}
	\vspace{-2pt}
	\caption{Illustration of the Attention driven adaptive denoising mechanism.}
	\label{fig:attention}
	\vspace{-12pt}
\end{figure}

To capture content-dependent importance across image regions, we introduce an attention-based adaptive denoising strategy that leverages the self-attention maps of a pretrained vision transformer as shown in Fig. \ref{fig:attention}. The input image is divided into $N$ non-overlapping patches $\{p_1, p_2, \dots, p_N\}$, and the semantic relevance of each patch is inferred using the average attention weights assigned to its corresponding token. Although the previous work in \cite{Adadiff} has explored adaptive denoising strategies, our method uniquely leverages self-attention maps from the pretrained model to guide the denoising, offering a more precise and content-aware denoising strategy. Furthermore, our method is straightforward to implement, requiring no additional training or policy learning, and thereby offering a more efficient and interpretable solution for adaptive denoising.

Given the attention matrix $A \in \mathbb{R}^{N \times N}$ from a selected transformer layer, the importance score $w_i$ for patch $p_i$ is computed as

\vspace{-8pt}
\begin{equation}
	\vspace{-2pt}
	w_i = \frac{1}{N} \sum_{j=1}^{N} A_{j,i},
\end{equation}

\noindent where $A_{j,i}$ denotes the attention weight from token $j$ to $i$, representing how much other patches attend to $p_i$.

The per-patch denoising step $n_i$ is linearly assigned based on normalized attention scores

\vspace{-5pt}
\begin{equation}
	\vspace{-2pt}
	n_i = n_{\min} + w_i^{\mathrm{norm}} \cdot (n_{\max} - n_{\min}),
\end{equation}

\noindent where $n_{\max}$, $n_{\min}$ are the maximum and minimum denoising steps and $w_i^{\mathrm{norm}} = \frac{w_i - \min w_i}{\max w_i - \min w_i}$ ensures $w_i^{\mathrm{norm}} \in [0,1]$.

During the reverse diffusion, a spatial mask is generated at each step to update only the regions with remaining denoising steps. This ensures an efficient computational efficiency while prioritizing semantically critical patches.

\vspace{-8pt}
\section{Numerical Experiments}
\label{sec:Experiments}
In this section, we evaluate the effectiveness of our designed system by comparing it with multiple baselines across various datasets. Our experimental results demonstrate the high performance and efficiency of the proposed model.

\vspace{-8pt}
\subsection{Experiments Setting}

The experiments were conducted on a server equipped with an AMD EPYC 7742 CPU and a NVIDIA A100-80GB GPU. The operating system is Linux, and CUDA 12.4 and PyTorch 2.6.0 are used for training and inference of the deep learning model.

The datasets utilized in our simulation experiments are ImageNet-256 and STL-10. ImageNet-256 is a downsized version of ImageNet, comprising 1.2 million training images resized to 256×256 pixels across 1,000 classes. STL-10 includes 5,000 labeled training images and 8,000 test images, each with a resolution of 96×96 pixels, spanning 10 classes, along with an additional 100,000 unlabeled images. These datasets are chosen to assess the performance of our method across various scales and complexities. By conducting tests on both datasets, we can demonstrate the robustness and versatility of our proposed approach under diverse conditions. In the simulation experiments, we aim to test and compare our system in both AWGN and Rayleigh channels. During the training and inference phases, we set training epochs to 50,000 and the batch size to 64 with hyperparameters $\beta=0.5$, $n_{\min}=100$, $n_{\max}=200$, $N=16$ and $T=1000$.

\vspace{-10pt}
\subsection{Performance Comparison}
\vspace{-4pt}

We compare our designed JSCNA-AD approach with other SemCom methods and traditional communication methods in two datasets. We examine channel transmission conditions with SNR ranging from 0 dB to 20 dB under AWGN and Rayleigh fading channels. The following methods are considered as baselines: the JPEG2000 encoding method with a low-density parity-check (LDPC) error-correcting code mechanism \cite{jpegldpc}; the Deep Joint Source-Channel Coding (DeepJSCC) system \cite{DeepJSCC}; and a transformer-based WITT SemCom system \cite{WITT} and JSCNA-AD system operates under ablation conditions without the adaptive denoising, utilizing  fixed steps of 150. The evaluation metrics we employ are PSNR and Multi-Scale Structural Similarity Index (MS-SSIM) performance under various channel conditions. These metrics assess the transmission results in terms of signal fidelity and image similarity.

\begin{figure}[ht]
	\vspace{-6pt}
	\centering
	\includegraphics[scale=0.158]{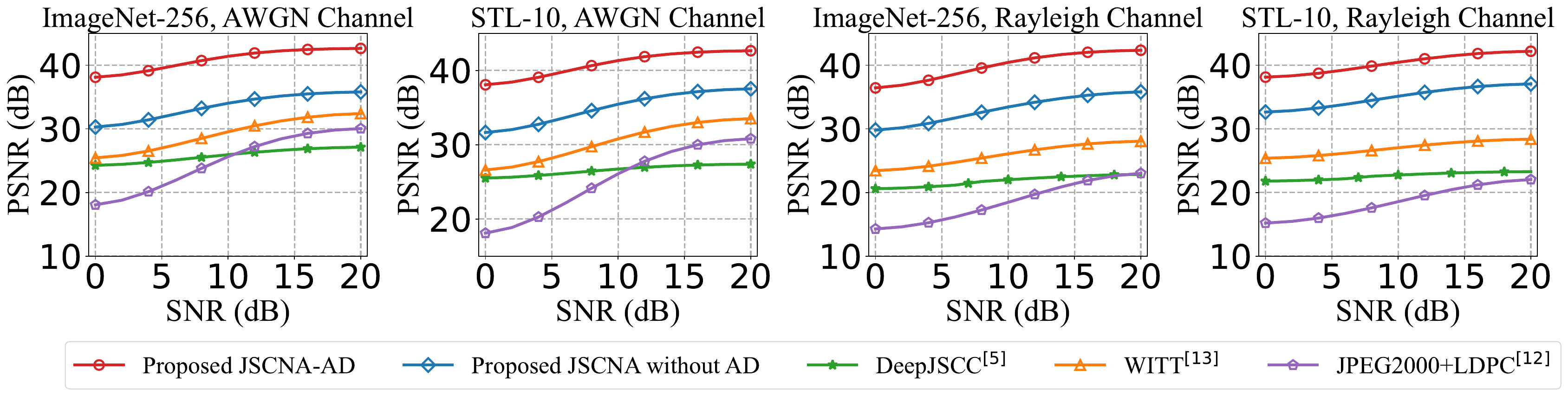}
	\caption{Comparison of the PSNR performance in ImageNet-256 and STL-10 datasets with AWGN and Rayleigh fading.}
	\label{fig:psnr}
	\vspace{-8pt}
\end{figure}

\begin{figure}[ht]
	\vspace{-8pt}
	\centering
	\includegraphics[scale=0.158]{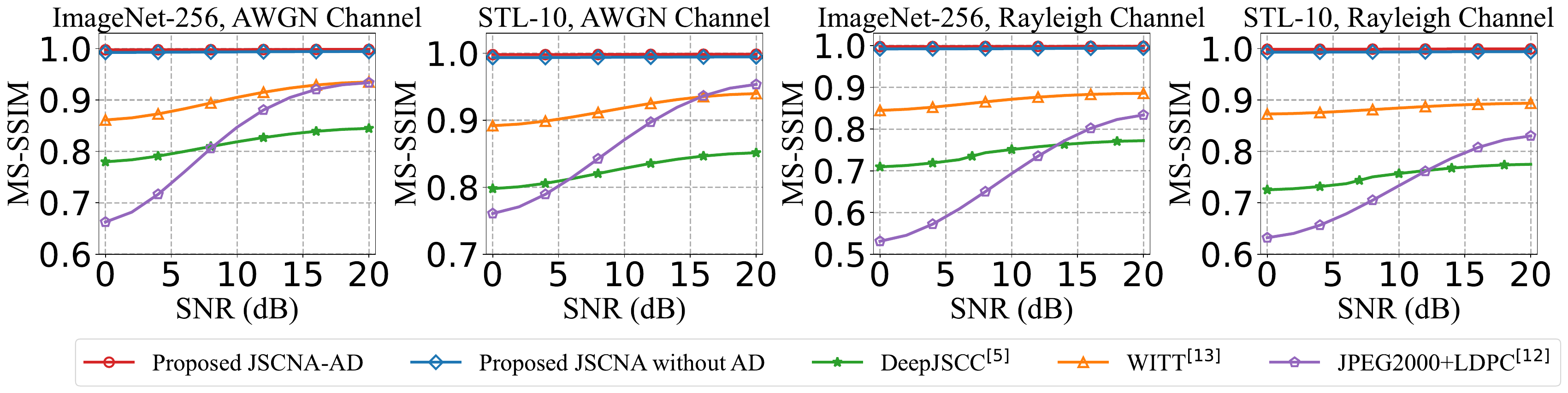}
	\caption{Comparison of the MS-SSIM performance in ImageNet-256 and STL-10 datasets with AWGN and Rayleigh fading.}
	\label{fig:msssim}
	\vspace{-6pt}
\end{figure}

Fig. \ref{fig:psnr} and Fig. \ref{fig:msssim} illustrates the PSNR and MS-SSIM performance of our proposed system compared to each baseline across two datasets under various channel conditions. For both datasets and under various channel conditions, the PSNR values of the JSCNA-AD system exceed 35 dB, and the MS-SSIM is very close to the maximum value of 1. The figure clearly demonstrates that our proposed JSCNA-AD system significantly outperforms other SemCom systems, such as WITT, as well as deep learning-based communication systems like DeepJSCC and traditional communication methods, including JPEG2000 with LDPC coding. Additionally, our system shows a marked improvement by comparing fixed and adaptive denoising steps, as evidenced by the ablation experiments. This highlights that the number of denoising steps determined by the attention-driven adaptive denoising effectively ensures transmission quality.


\begin{table}[t]
	\centering
	\captionsetup{justification=centering} 
	\renewcommand{\arraystretch}{1.5}
	\caption{\centering COMPARISON OF COMPUTATIONAL COMPLEXITY FOR TRANSMISSION OF DIFFERENT MODELS}
	\centering
	\scriptsize
	\begin{tabular}{|@{\hspace{3pt}}c@{\hspace{3pt}}|@{\hspace{3pt}}c@{\hspace{3pt}}|@{\hspace{3pt}}c@{\hspace{3pt}}|@{\hspace{3pt}}c@{\hspace{3pt}}|@{\hspace{3pt}}c@{\hspace{3pt}}|}
		\hline
		\textbf{Method} & \textbf{Parameters} & \textbf{FLOPs} & \textbf{Inference Time} & \textbf{PSNR}\\
		\hline
		Proposed JSCNA-AD & 7.55M  & 154.37G    & 109 ms & 36.41 dB \\
		\hline
		JSCNA without AD  & 7.55M  & 148.58G    & 469 ms & 29.98 dB \\
		\hline
		WITT              & 28.20M & 198.17G    & 116 ms & 23.10 dB \\
		\hline
		JPEG2000+LDPC     & N/A    & 0.05G  & 605 ms & 17.22 dB \\ 
		\hline
		DeepJSCC          & 0.19M  & 0.01G   & 238 ms & 21.86 dB \\
		\hline
	\end{tabular}
	\label{table:comp}
\end{table}

We present statistics on the computational complexity of several algorithms, along with the inference time and performance at the receiver over a Rayleigh channel with an SNR of 4 dB in ImageNet-256 dataset, as shown in Table \ref{table:comp}. The table indicates that the JSCNA-AD system we designed has fewer model parameters and requires fewer floating-point calculations compared to other SemCom systems. Furthermore, the inference time is shorter than that of other methods, and the transmission performance is significantly superior to that of any other algorithm.

\vspace{-10pt}
\section{Conclusion} 
\label{sec:conclusion}

In this paper, we propose a novel diffusion-based JSCNA-AD framework that integrates channel noise addition, channel-aware denoising, and the attention-guided adaptive denoising strategy. By modeling the communication process through the lens of probabilistic diffusion dynamics, our system achieves robust semantic reconstruction even under severe channel degradation. The proposed attention mechanism enables the transmitter to proactively simulate denoising outcomes and optimize transmission strategies, thereby striking a practical balance between latency and fidelity. Extensive experiments demonstrate the superiority of our approach over conventional SemCom baselines across diverse noise and fading conditions.

\ifCLASSOPTIONcaptionsoff
\newpage
\fi

\bibliographystyle{IEEEtran}
\vspace{-7pt}
\bibliography{refer.bib}

\begin{thebibliography}{10}
\providecommand{\url}[1]{#1}
\csname url@samestyle\endcsname
\providecommand{\newblock}{\relax}
\providecommand{\bibinfo}[2]{#2}
\providecommand{\BIBentrySTDinterwordspacing}{\spaceskip=0pt\relax}
\providecommand{\BIBentryALTinterwordstretchfactor}{4}
\providecommand{\BIBentryALTinterwordspacing}{\spaceskip=\fontdimen2\font plus
\BIBentryALTinterwordstretchfactor\fontdimen3\font minus
  \fontdimen4\font\relax}
\providecommand{\BIBforeignlanguage}[2]{{%
\expandafter\ifx\csname l@#1\endcsname\relax
\typeout{** WARNING: IEEEtran.bst: No hyphenation pattern has been}%
\typeout{** loaded for the language `#1'. Using the pattern for}%
\typeout{** the default language instead.}%
\else
\language=\csname l@#1\endcsname
\fi
#2}}
\providecommand{\BIBdecl}{\relax}
\BIBdecl

\bibitem{AComplete}
G.~Qiu, G.~Tang, C.~Li, L.~Luo, D.~Guo, and Y.~Shen, ``A complete and
  comprehensive semantic perception of mobile traveling for mobile
  communication services,'' \emph{IEEE Internet Things J.}, vol.~11, no.~3, pp.
  5467--5490, Feb. 2024.

\bibitem{SCcarrier}
Z.~Yan and D.~Li, ``Semantic communications for digital signals via carrier
  images,'' \emph{IEEE Wireless Commun. Lett.}, vol.~14, no.~6, pp. 1816--1820,
  Jun. 2025.

\bibitem{DeepJSCC}
E.~Bourtsoulatze, D.~Burth~Kurka, and D.~Gündüz, ``Deep joint source-channel
  coding for wireless image transmission,'' \emph{IEEE Trans. Cognit. Commun.
  Networking}, vol.~5, no.~3, pp. 567--579, Sep. 2019.

\bibitem{NLJSCNSC}
X.~Yu, D.~Li, N.~Zhang, and X.~Shen, ``A novel lightweight joint source-channel
  coding design in semantic communications,'' \emph{IEEE Internet Things J.},
  vol.~12, no.~11, pp. 18\,447--18\,450, Jun. 2025.

\bibitem{DDPM}
J.~Ho, A.~Jain, and P.~Abbeel, ``Denoising diffusion probabilistic models,'' in
  \emph{Proc. Adv. Neural Inf. Process. Syst. (NeurIPS)}, vol.~33, 2020, pp.
  6840--6851.

\bibitem{YOLO}
B.~Du, H.~Du, H.~Liu, D.~Niyato, P.~Xin, J.~Yu, M.~Qi, and Y.~Tang,
  ``Yolo-based semantic communication with generative ai-aided resource
  allocation for digital twins construction,'' \emph{IEEE Internet Things J.},
  vol.~11, no.~5, pp. 7664--7678, Mar. 2024.

\bibitem{SelectSC}
C.~Liang, D.~Li, Z.~Lin, and H.~Cao, ``Selection-based image generation for
  semantic communication systems,'' \emph{IEEE Commun. Lett.}, vol.~28, no.~1,
  pp. 34--38, Dec. 2024.

\bibitem{Adadiff}
Y.~Fan, C.~Liu, N.~Yin, C.~Gao, and X.~Qian, ``Adadiffsr: Adaptive region-aware
  dynamic acceleration diffusion model for real-world image super-resolution,''
  in \emph{Proc. Eur. Conf. Comput. Vis. (ECCV) 2024}, Dec. 2024, pp. 396--413.

\bibitem{jpegldpc}
L.~Pu, Z.~Wu, A.~Bilgin, M.~W. Marcellin, and B.~Vasic, ``Ldpc-based iterative
  joint source-channel decoding for jpeg2000,'' \emph{IEEE Trans. Image
  Process.}, vol.~16, no.~2, pp. 577--581, Jan. 2007.

\bibitem{WITT}
K.~Yang, S.~Wang, J.~Dai, K.~Tan, K.~Niu, and P.~Zhang, ``Witt: A wireless
  image transmission transformer for semantic communications,'' in \emph{Proc.
  IEEE Int. Conf. Acoust. Speech Signal Process. (ICASSP)}, 2023, pp. 1--5.

\end{thebibliography}

\end{document}